\documentclass[twocolumn]{el-author}
\usepackage{subscript}

\usepackage{tabularx} 
\usepackage{algpseudocode}
\usepackage{algorithm}

\usepackage{color} 
\usepackage{url}
\usepackage{hyperref}
\usepackage{breakurl}

\algrenewcommand\algorithmicindent{1.0em}
\algtext*{EndIf}
\algtext*{EndFor}
\algtext*{EndWhile}

\newcommand{\ua}{\uparrow}
\newcommand{\nc}{\newcommand}
\nc{\da}{\downarrow} \nc{\hc}{\hat{c}} \nc{\hS}{\hat{S}}
\nc{\bra}{\langle} \nc{\ket}{\rangle} \nc{\eq}{equation (\ref}
\nc{\h}{\hat} \nc{\hT}{\h{T}}\nc{\be}{\begin{eqnarray}}
\nc{\ee}{\end{eqnarray}}\nc{\rd}{\textrm{d}}\nc{\e}{eqnarray}\nc{\hR}{\hat{R}}\nc{\Tr}{\mathrm{Tr}}
\nc{\tS}{\tilde{S}}\nc{\tr}{\mathrm{tr}}\nc{\8}{\infty}\nc{\lgs}{\bra\ua,\phi|}\nc{\rgs}{|\ua,\phi\ket}
\nc{\hU}{\hat{U}}\nc{\lfs}{\bra\phi|}\nc{\rfs}{|\phi\ket}\nc{\hZ}{\hat{Z}}\nc{\hd}{\hat{d}}\nc{\mD}{\mathcal{D}}
\nc{\bd}{\bar{d}}\nc{\bc}{\bar{c}}\nc{\mc}{\mathcal}\nc{\ea}{eqnarray}\nc{\mG}{\mathcal{G}}\nc{\bce}{\begin{center}}
\nc{\ece}{\end{center}}
\date{23rd November 2015}

\begin{document}
\title{Reordering GPU Kernel Launches to Enable Efficient Concurrent Execution}
\author{Teng Li, Vikram K. Narayana and Tarek El-Ghazawi}
\abstract{Contemporary GPUs allow concurrent execution of small computational kernels in order to prevent idling of GPU resources. 
Despite the potential concurrency
between independent kernels, the order in which kernels are issued to the GPU will significantly influence the application performance.  A
technique for deriving suitable kernel launch orders is therefore presented, with the aim of reducing the total execution time.  
Experimental results indicate that the proposed method yields solutions that are well above the 90 percentile mark in the design
space of all possible permutations of the kernel launch sequences. }

\maketitle
\section{Introduction}

Graphics processing units (GPU) have experienced widespread adoption in the scientific computing community as application accelerators.  
Programmers encapsulate parts of their application as compute kernels for execution on the GPU co-processor, by using 
language extensions such as NVIDIA's CUDA~\cite{m_cuda6}. Frequently, these compute kernels cannot completely utilize the GPU resources.  Vendors have 
therefore introduced features of concurrent execution of kernels, thereby enabling increased resource utilization and an overall reduction in the 
GPU execution time. For NVIDIA GPUs, concurrency is 
achieved by queueing independent kernels into separate CUDA streams.  
When a limited number of streams are deployed, it is a well-known fact that the practically achieved parallelism is affected by the order in which kernels
are enqueued into their respective streams, due to false dependencies arising from hardware and software limitations~\cite{webinar12}.  To avoid these false dependencies, 
users can dedicate one stream for every kernel, as long as the kernels are independent.  However, researchers have overlooked the fact that even in this case,
the order in which the streams are initiated can significantly influence the concurrency and thus the total execution time.  For instance, a recent study~\cite{lu2012gpu} 
reported that the effect of kernel launch order on the total execution time is insignificant; however, their conclusion was erroneous because it was based on 
identical kernels differing only in the number of thread blocks within each experiment.  As we shall see shortly, ordering does not matter for that case.  
Only very recently, 
Pai \emph{et al}~\cite{pai2013improving} identified this issue of ``non-commutative concurrency'' for GPUs; nevertheless, their solution follows a different approach 
through source to source transformation of kernels into elastic versions, whereas we propose the reordering of kernel launch orders without any kernel modification. Li \emph{et al}~\cite{cf14, icpads15, c_gpusch} also proposed several power/energy/performance-aware scheduding techniques for concurrent GPU kernel executions. The work was primarily to support efficient GPU sharing \cite{c_vgpu, cf12, computers} by improving the overall GPU resource utilization through efficient kernel scheduling algorithms.





\begin{table}[b]
  \processtable{GPU and Kernel Parameters\textsuperscript{*} \label{tb:para}}
	\centering
	{\begin{tabular}{l|l|l|l}
		\hline 
		\emph{N\textsubscript{SM}} & \# of SMs in the GPU & \emph{N\textsubscript{reg\_SM}} & \# of registers per SM \\
		\hline
		\emph{N\textsubscript{shm\_SM}} & Shared mem size per SM & \emph{N\textsubscript{warp\_SM}} & Max \# of warps per SM \\
		\hline
		\emph{N\textsubscript{blk\_SM}} & Max \# of blocks per SM & \emph{R\textsubscript{B}} & Balanced Inst/Mem ratio\\
		\hline
		\hline
		\emph{N\textsubscript{inst\_i}} & \# of inst for kernel \emph{i}& \emph{N\textsubscript{reg\_i}} & \# of registers for kernel \emph{i}  \\
		\hline
		\emph{N\textsubscript{shm\_i}} & Shared mem size for kernel \emph{i} & \emph{N\textsubscript{warp\_i}} & \# of warps for kernel \emph{i}\\
		\hline
		 \emph{N\textsubscript{tblk\_i}} & \# of blocks for kernel \emph{i} & \emph{R\textsubscript{i}} & Inst/Mem ratio for kernel \emph{i}\\
		\hline

	\end{tabular}}{\scriptsize \textsuperscript{*}The first three rows are constant for a GPU, whereas the remainings are kernel-specific.}
\end{table} 
\section{Fundamental Concept of Reordering}
 GPU cores, or streaming processors (SP), are organized into groups known as
 streaming multiprocessors (SM).  Each SM executes one or more thread blocks. 
 When there are several kernels ready for execution, all thread blocks from the earliest issued kernel are first allocated to the SMs, followed by thread blocks 
 from the next issued kernel~\cite{pai2013improving}. If the total number of thread blocks does not exceed \emph{N\textsubscript{SM}}, kernels do not share any SM. 
 In this case the launch order does not have an impact on the total execution time.  
 On the other hand, with a larger number of thread blocks, multiple thread blocks from one or more kernels will need to 
 share an SM.  For instance, if 
 there are \emph{2N\textsubscript{SM}} thread blocks in total, each SM will be assigned two thread blocks.  
 In general, additional thread blocks are mapped to SMs in a round-robin fashion, until any one of the SM resource limitations is met: 
 \emph{N\textsubscript{reg\_SM}}, \emph{N\textsubscript{shm\_SM}},
 \emph{N\textsubscript{warp\_SM}} and \emph{N\textsubscript{blk\_SM}}, as defined in Table \ref{tb:para}. 
  When a kernel consumes just one of the SM resources and leaves other resources underutilized, it prevents additional thread blocks from 
  being assigned to the SM, and those thread blocks are relegated to the next \emph{execution round}.  Therefore, thread blocks from 
  a set of kernels are split into multiple \emph{execution rounds}, which are sequentially executed one after the other.  Concurrency within 
  each round depends on how much resources are utilized; an ill-suited launch order can result in just one of the SM resources being heavily
  utilized thereby limiting the number of concurrent kernels within an \emph{execution round}, which can lead to a reduced performance.  Our goal is thus to obtain a launch order that maximizes the utilization of all SM 
  resources within an \emph{execution round}.

\section{Scope and Applicability} 
Reordering is useful only when the total number of thread blocks exceeds \emph{N\textsubscript{SM}}, which is normally the case.  
Even in this case, 
if the kernels are identical and differ only in the number of thread blocks, the composition of each \emph{execution round} and the number of \emph{rounds} is the same regardless of the order, because a thread block cannot split across SMs.  In this specific case, the order will not matter.   
%
Additionally, even if the kernels are non-identical, it might so happen that the thread block of every kernel is resource-heavy and the SM can accommodate only 
one thread block at a time; in this case too, the order will not impact the performance.  Our work thus covers only the most common cases.  

%

\section{Balancing Compute \& Memory Accesses} 
Apart from resource limitations, multi-kernel execution performance is affected by the balance of compute and memory accesses.  As indicated by NVIDIA,
even for a single kernel there exists a suitable target value \emph{R\textsubscript{B}} for the balanced instructions/bytes ratio, and we use the same concept for multiple kernels. 
For each \emph{execution round}, we aim to achieve a combined instructions/bytes ratio \emph{R\textsubscript{comb}} that is as close to \emph{R\textsubscript{B}} as possible.  This translates
to having memory-bound kernels launching in close proximity to compute-bound kernels. 
Using CUDA profiler data from the individual kernels, we can compute 
\emph{R\textsubscript{comb}} = \emph{total \# of instructions} / 4*(\emph{total \# of global stores + total \# of L1 cache global load misses}). 

\begin{algorithm}
  	\scriptsize
	\caption{Concurrent Kernel Launch Order Algorithm}
	\label{alg:symbiosis}
	\begin{algorithmic}[5]
		\renewcommand{\algorithmicrequire}{\textbf{Input:}}
		\renewcommand{\algorithmicensure}{\textbf{Output:}}
		\Require the set of \emph{N\textsubscript{knl}} kernels (\emph{K}) with profiling results (\emph{PR}): \emph{N\textsubscript{tblk\_i}},\emph{N\textsubscript{reg\_i}},\emph{N\textsubscript{shm\_i}},\emph{N\textsubscript{warp\_i}},\emph{R\textsubscript{i}}
		\State Denote \emph{Rd\textsubscript{r}} to be the set storing kernel order within \emph{execution round r}; r=0 
		\State ScoreMatrix[][]=ScoreGen(\emph{K}, \emph{K}, \emph{PR})
		\While{\emph{K} != null}
		\State r++ \Comment{Counting towards the next \emph{execution round}}		
		\State Within \emph{K}, find kernel \emph{K\textsubscript{a}},\emph{K\textsubscript{b}} with highest score in ScoreMatrix[][]
		\State Push \emph{K\textsubscript{a}},\emph{K\textsubscript{b}} into \emph{Rd\textsubscript{r}} (using decreasing order of \emph{N\textsubscript{shm\_a}}, \emph{N\textsubscript{shm\_b}}) and remove from \emph{K} 
		\State \emph{K\textsubscript{comb}}=ProfileCombine(\emph{K\textsubscript{a}},\emph{K\textsubscript{b}})

		\For {All kernels \emph{K\textsubscript{r}} (from \emph{K}) whose resource can fit within \emph{Rd\textsubscript{r}}}
		\State ScoreVec[]=ScoreGen(\emph{K\textsubscript{comb}}, \emph{K\textsubscript{r}}, \emph{PR})
		\State Push \emph{K\textsubscript{c}} with the highest score in ScoreVec[] into \emph{Rd\textsubscript{r}} (Sort by \emph{N\textsubscript{shm\_c}}, \emph{N\textsubscript{shm\_comb}}) 
		\State \emph{K\textsubscript{comb}}=ProfileCombine(\emph{K\textsubscript{comb}},\emph{K\textsubscript{c}}) and remove \emph{K\textsubscript{c}} from \emph{K}
		\EndFor
		\EndWhile
		\Ensure Kernel launch order from \emph{Rd\textsubscript{1}} to \emph{Rd\textsubscript{r}} 
		\vspace{-4.5pt}
		\\\dotfill
		\vspace{-1.5pt}
		\Function{ScoreGen}{\emph{K\textsubscript{M}}, \emph{K\textsubscript{N}}, \emph{PR}} \Comment{\emph{K\textsubscript{M}} \& \emph{K\textsubscript{N}} are two kernel sets}
		\For {All kernels \emph{K\textsubscript{i}} within \emph{K\textsubscript{M}}}
		\For {All kernels \emph{K\textsubscript{j}} within \emph{K\textsubscript{N}}}
		\If{\emph{K\textsubscript{i}} and \emph{K\textsubscript{j}} cannot fit within an \emph{execution round}} 
		S[i][j] = 0
		\Else


		\State S[i][j] += \textbf{max}\{(\emph{N\textsubscript{shm\_SM}}-\emph{N\textsubscript{shm\_i}}-\emph{N\textsubscript{shm\_j}})/\emph{N\textsubscript{shm\_SM}}, 0\}


		\State S[i][j] += \textbf{max}\{(\emph{N\textsubscript{reg\_SM}}-\emph{N\textsubscript{reg\_i}}-\emph{N\textsubscript{reg\_j}})/\emph{N\textsubscript{reg\_SM}}, 0\}


		\State S[i][j] += \textbf{max}\{(\emph{N\textsubscript{warp\_SM}}-\emph{N\textsubscript{warp\_i}}-\emph{N\textsubscript{warp\_j}})/\emph{N\textsubscript{warp\_SM}}, 0\}

		\If{\emph{R\textsubscript{i}}$\le$\emph{R\textsubscript{B}}$\le$\emph{R\textsubscript{j}} \textbf{or} \emph{R\textsubscript{j}}$\le$\emph{R\textsubscript{B}}$\le$\emph{R\textsubscript{i}}}
		\State S[i][j]+= \textbf{max}\{1-(|\emph{R\textsubscript{comb(i,j)}}-\emph{R\textsubscript{B}}|/\emph{R\textsubscript{B}}), 0\} \Comment{\emph{R\textsubscript{comb(i,j)}} is the combined ratio}
		\EndIf
		\EndIf
		\EndFor
		\EndFor
		\Return{S[][]} 
		\EndFunction
		\vspace{-4.5pt}
		\\\dotfill
		\vspace{-1.5pt}
		\Function{ProfileCombine}{\emph{K\textsubscript{a}}, \emph{K\textsubscript{b}}}
		\State \emph{N\textsubscript{shm\_comb}}=\emph{N\textsubscript{shm\_a}}+\emph{N\textsubscript{shm\_b}};  \emph{N\textsubscript{reg\_comb}}=\emph{N\textsubscript{reg\_a}}+\emph{N\textsubscript{reg\_b}};  \emph{N\textsubscript{warp\_comb}}=\emph{N\textsubscript{warp\_a}}+\emph{N\textsubscript{warp\_b}};
		\State \emph{N\textsubscript{tblk\_comb}}=\emph{N\textsubscript{tblk\_a}}+\emph{N\textsubscript{tblk\_b}};  \emph{R\textsubscript{comb}}=\emph{R\textsubscript{comb(a,b)}}=(\emph{N\textsubscript{inst\_a}}+\emph{N\textsubscript{inst\_b}})/(\emph{N\textsubscript{inst\_a}}/\emph{R\textsubscript{a}}+\emph{N\textsubscript{inst\_b}}/\emph{R\textsubscript{b}})
		\State \Return{\emph{K\textsubscript{comb}}} \Comment{Virtual ``kernel'' with combined profile }
		\EndFunction
	\end{algorithmic}
\end{algorithm}
\section{Proposed Algorithm}
Considering both factors - SM resources and balanced compute/memory - we propose and implement (using C) a greedy algorithm for scheduling GPU kernels. The basic idea is to select the kernel launch order such that the number of kernels within an \emph{execution round} is maximized, and the SM resources are progressively utilized in a balanced manner
as kernels arrive.
Selection of kernels is made sequentially based on a computed score. \emph{ScoreGen(K\textsubscript{X}, K\textsubscript{Y})} computes the score between every kernel pair 
taken from the set \emph{K\textsubscript{X}} and \emph{K\textsubscript{Y}} respectively.  The resultant score matrix is two dimensional or one dimensional
depending on the input dimensions.  For every kernel pair, the resulting SM resources that remain available add to the score, lines 18-20 in Algorithm~\ref{alg:symbiosis} (see Table~\ref{tb:para} 
for symbol definitions).  Kernel pairs that result in a balanced (and lower) usage of all three resources result in the highest score, allowing more 
subsequent kernels to co-execute within the \emph{execution round}.  Similarly, a higher score is provided if the resulting instructions/bytes ratio for the 
\emph{execution round} is closer to the target value \emph{R\textsubscript{B}}, line 22 in Algorithm~\ref{alg:symbiosis}.  Note that the conditional statement in line 21 ensures that a score is added only if the kernels under consideration are of opposing type, i.e., compute-bound vs memory-bound, because \emph{R\textsubscript{B}} is deemed to be the ratio
for an ideal, balanced kernel that is neither compute-bound nor memory-bound.   

 For each \emph{execution round} \emph{{r}}, a pair of kernels with the highest score is selected and inserted into the round, denoted
by the set \emph{Rd\textsubscript{r}}. The inserted pair's order is sorted decreasingly by shared memory usage since this allows kernels with more \emph{N\textsubscript{shm\_i}} to finish faster, and thus release \emph{N\textsubscript{shm\_i}} sooner. The kernel pair is virtually combined by profile into a virtual kernel \emph{K\textsubscript{comb}} with function \emph{ProfileCombine()} so that the overall resource of current \emph{Rd\textsubscript{r}} can be taken into account when choosing the next kernel for the \emph{execution round}. Kernels continue to 
be incorporated into the round \emph{{r}} as long as resources permit until a new round \emph{{r+1}} needs to be opened.


\begin{figure}[t]
      \centering
      \includegraphics[width=1\linewidth]{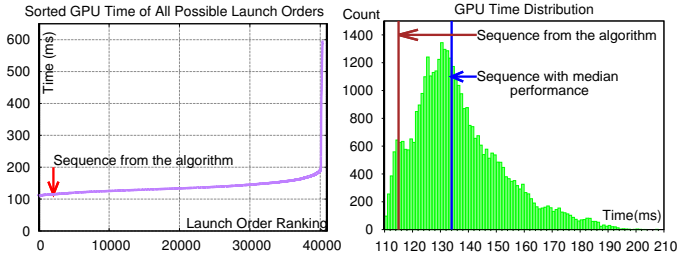}
      \caption{Ranking and Distribution of GPU Execution Time in the Launch Order Permutation Space for EpBsEsSw-8}
      \label{fig:4kernels}
      
  \end{figure} 

\begin{table}[t]
  \processtable{Experiment Parameters\label{tb:bench}}
   \centering
   {\begin{tabular}{l|p{2.2cm}|p{4.0cm}}
   \hline
		\emph{Experiment} & Constant Parameters & Variables Across Kernels \\
   \hline
   \emph{EP-6-shm} & \makebox[2.2cm][l]{\emph{R\textsubscript{i}}=3.11,} \mbox{16\textsubscript{Grid Size} x 128\textsubscript{Block Size}} & \emph{N\textsubscript{shm\_i}} = 8K, 16K, 24K, 32K, 40K, 48K \\ 
	\hline
	\emph{EP-6-grid} & \makebox[2.2cm][l]{\emph{R\textsubscript{i}}=3.11, \emph{N\textsubscript{shm\_i}} = 0,} \mbox{128\textsubscript{Block Size}}  & \mbox{\emph{N\textsubscript{warp\_i}} = 4, 8, 12, 16, 20, 24} \mbox{(\textsubscript{Grid Size} = 16, 32, 48, 64, 80, 96)}\\
	\hline
	\emph{BS-6-blk} & \makebox[2.2cm][l]{\emph{R\textsubscript{i}}=11.1, \emph{N\textsubscript{shm\_i}} = 0,} \mbox{32\textsubscript{Grid Size}} & \mbox{\emph{N\textsubscript{warp\_i}} = 4, 8, 12, 16, 32, 64} \mbox{(\textsubscript{Block Size} = 64, 128, 256, 512, 768, 1024)}\\
	\hline
        \emph{EpBs-6} & \emph{N\textsubscript{shm\_i}} = 0 & \makebox[4.0cm][l]{3 EP kernels w/ \emph{N\textsubscript{warp\_i}}=4, \emph{R\textsubscript{i}}=3.11} 3 BS kernels w/ \emph{N\textsubscript{warp\_i}}=12, \emph{R\textsubscript{i}}=11.1 \\ 
	\hline
	\emph{EpBs-6-shm} & \makebox[2.2cm][c]{\textemdash} & \makebox[4.0cm][l]{3 EP w/ \emph{N\textsubscript{warp\_i}}=4, \emph{N\textsubscript{shm\_i}}=16K,24K,48K} \makebox[4.0cm][l]{3 BS w/ \emph{N\textsubscript{warp\_i}}=12, \emph{N\textsubscript{shm\_i}}=16K,24K,48K}\\
	\hline   
	\emph{EpBsEsSw-8} & \makebox[2.2cm][c]{\textemdash}  & EP, BS, ES and SW kernels, 2 each\\    
	\hline   
\end{tabular}}{}
\end{table}

  \begin{table}[t]
	\processtable{Experimental Results (GPU execution time) and Comparisons\label{tb:results}}
	\centering
	{\begin{tabular}{@{}l@{}|p{0.65cm}|p{0.65cm}|p{0.85cm}|p{1.0cm}@{}|p{1.1cm}@{}|p{1.3cm}@{}}
		\hline 
		\emph{Experiment} & Optimal (ms) & Worst (ms) & Algorithm (ms) & Percentile rank & Speedup over worst & Deviation from optimal \\
		\hline
		\emph{EP-6-shm} & 140.46 & 249.15 & 146.38 & 91.5\% & 1.702 & 4.21\% \\
		\hline
		\emph{EP-6-grid} & 123.39 & 156.03 & 123.45 & 96.3\% & 1.264 & 0.049\%\\
		\hline
		\emph{BS-6-blk} & 699.29 & 1699.04 & 702.29 & 96.5\% & 2.419 & 0.43\%\\
		\hline
		\emph{EpBs-6} & 100.03 & 167.47 & 100.20 & 96.1\% & 1.671 & 0.17\%\\
		\hline
		\emph{EpBs-6-shm} & 251.90 & 311.79 & 251.95 & 99.4\% & 1.238 & 0.02\%\\
		\hline
		\emph{EpBsEsSw-8} & 109.21& 597.43 & 115.23 & 94.8\% & 5.185 & 5.51\%\\
		\hline
	\end{tabular}}{}
\end{table} 

\section{Experimental Results}
The experimental platform is a GPU computing node with dual Intel Xeon X5570 CPUs and an NVIDIA GTX580 GPU (16 SMs, \emph{R\textsubscript{B}}={4.11}, \emph{N\textsubscript{reg\_SM}}=32K, \emph{N\textsubscript{warp\_SM}}=48, \emph{N\textsubscript{shm\_SM}}=48K, \emph{N\textsubscript{blk\_SM}}=8). All benchmark results are collected under Ubuntu 11.10 with CUDA 5.0 while \emph{N\textsubscript{tblk\_i}}, \emph{N\textsubscript{reg\_i}}, \emph{N\textsubscript{shm\_i}}, \emph{N\textsubscript{warp\_i}} and \emph{R\textsubscript{i}} are analyzed using CUDA profiler. Our experiments evaluate the concurrent execution time of all possible kernel orderings (all permutations) and compare the performance of the kernel ordering given by the algorithm with the optimal (best) result. The percentile rank among all permutations, the speedup over the worst case and the deviation from the optimal result for the algorithm results are also presented, as shown in Table~\ref{tb:results}. To demonstrate the effectiveness of our algorithm on different resource metrics, we initially conduct six experiments, each of which consists of six concurrent kernels. We use NAS Parallel Benchmarks (NPB) kernel EP (M=24) (\emph{R\textsubscript{ep}=3.11} < \emph{R\textsubscript{B}}) \cite{CPE:CPE1860} and the European option pricing benchmark BlackScholes (BS) (4M options) (\emph{R\textsubscript{bs}=11.1} > \emph{R\textsubscript{B}}) as two applications to represent memory-bound and compute-bound respectively. The experiment parameters are summarized in Table~\ref{tb:bench}.  \emph{EP-6-shm} consists of six EP kernels that varies only the shared memory usage, whereas \emph{EP-6-grid} varies only the warp usage by changing just the kernel grid size. 
 The experiment \emph{BS-6-blk} again varies only the warps, but this time by changing the block size alone. Thus, \emph{EP-6-grid} and \emph{BS-6-blk} both demonstrate the effectiveness of algorithm on varied \emph{N\textsubscript{warp\_i}}, as shown in Table~\ref{tb:results}. The next experiment, \emph{EpBs-6} tests the same but with two different kernels with varied Inst/Mem ratios (\emph{R\textsubscript{i}}).  The effect of varying the shared memory is further factored in by running the \emph{EpBs-6-shm} experiment.  
From the comparison in Table \ref{tb:results}, all the six experiments with specific variation in resource metrics prove that the kernel launch order from the algorithm provides close-to-optimal results. We further conduct a more general experiment with four applications from different fields: the Electrostatics (ES) algorithm (40K atoms) from Visual Molecular Dynamics, Smith Waterman(SW) algorithm plus BS and EP. The experiment \emph{EpBsEsSw-8} is composed of 2 kernels of each application with a total of 8 kernels. With 4 different applications, kernels are varied with each other for all 
\emph{N\textsubscript{tblk\_i}}, \emph{N\textsubscript{reg\_i}}, \emph{N\textsubscript{shm\_i}}, \emph{N\textsubscript{warp\_i}}, \emph{R\textsubscript{i}} metrics. Fig.\ref{fig:4kernels} demonstrates the performance ranking of all possible kernel orderings for \emph{EpBsEsSw-8} while showing the near-optimal algorithm results with a percentile ranking of 94.8\%. It also shows the time distribution of all 40,320 permutations for \emph{EsBsEsSw-8}. By comparing the median sequence against the one from the algorithm, we demonstrate that our algorithm has 50\% of the probability to provide a minimum 16.1\% performance gain over a random order choice, and further up to 5.185 speedup over the worst case. 


\ack{This work was supported in part by the I/UCRC Program of the NSF under Grant Nos. IIP-1161014 and IIP-1230815.}

\vskip5pt

\noindent Teng Li, Vikram K. Narayana and Tarek El-Ghazawi (\textit{Department of Electrical and Computer Engineering, The George Washington University, 801 22nd St NW, Washington, DC, 20052, United States})
\vskip3pt

\noindent E-mail: \{tengli, tarek\}@gwu.edu; vikramkn@ieee.org

\end{document}